\documentclass{article}
\pdfoutput=1
\usepackage{arxiv}

\usepackage{pgfplots}
\pgfplotsset{compat=1.18}
\usepackage{placeins}
\usepackage{pgfplots}
\pgfplotsset{compat=newest}
\usepackage{xcolor}
\usepackage{multirow}
\usepackage{url}
\usepackage{booktabs}
\usepackage{nicefrac}       
\usepackage{microtype}      
\usepackage{lipsum}
\usepackage{graphicx}
\usepackage{amsmath}
\usepackage{amssymb}
\usepackage{mathtools}
\usepackage{amsthm}
\usepackage{algorithm}
\usepackage{algpseudocodex}
\usepackage{hyperref}
\usepackage{xr}
\usepackage{ulem}
\usepackage{placeins}

\makeatletter
\newcommand{\ALOOP}[1]{\ALC@it\algorithmicloop\ #1%
  \begin{ALC@loop}}
\newcommand{\ENDALOOP}{\end{ALC@loop}\ALC@it\algorithmicendloop}

\makeatother

\DeclareMathOperator*{\argmax}{arg\,max}

\def\musals{MuSAlS}
\begin{document}


\title{MuSAlS: A Fast Multiple Sequence Alignment Approach Using Hierarchical Clustering}

\author{
    Emily G. Light \\
    Department of Computer Science and Statistics \\
    University of Rhode Island \\
    Kingston, RI \\
    \And
    Morgan E. Prior \\
    Department of Computer Science \\
    Tufts University \\
    Medford, MA \\
    \And
    Noah M. Daniels \\
    Department of Computer Science and Statistics\\
    University of Rhode Island\\
    Kingston, RI\\
    \texttt{noah\_daniels@uri.edu} \\
    \And
    Najib Ishaq \\
    Department of Computer Science and Statistics\\
    University of Rhode Island\\
    Kingston, RI\\
    \texttt{najib\_ishaq@zoho.com}
}

\maketitle



\begin{abstract}Motivation: The multiple sequence alignment (MSA) problem has been extensively studied, with numerous approaches developed over recent years. With the rapid growth of sequence data, there is an increasing need for fast and accurate MSA tools that scale effectively to large datasets. Building on our previous work on CLAM, we are able to use exact dynamic programming (Needleman-Wunsch) while scaling to large datasets. We introduce MuSAlS (Multiple Sequence Alignment at Scale), a fast and scalable de novo MSA aligner. MuSAlS uses hierarchical clustering to construct a guide tree based on the Levenshtein distance metric, enabling efficient and accurate alignment through a bottom-up approach. Results: MuSAlS achieves competitive accuracy compared to state-of-the-art methods while significantly improving runtime performance. This makes it a valuable tool for researchers analyzing large-scale genomic and metagenomic datasets, addressing the growing demand for scalable bioinformatics solutions. Availability and Implementation: MuSAlS is implemented in the Rust programming language, and available at https://github.com/URI-ABD/clam.
\end{abstract}



\maketitle

\section{Introduction}
\label{sec:introduction}

The multiple sequence alignment (MSA) problem has been extensively studied, leading to the development of various approaches over the years. With the rapid growth of sequence data, the demand for faster and more accurate MSA methods has increased dramatically. The applications of MSA in molecular biology, computational biology, and bioinformatics make it an essential technique in these research fields. As Next-Generation Sequencing (NGS) techniques continue to advance, the speed of data generation has also increased. A fast and accurate MSA approach would have far-reaching implications in the field of biology, allowing researchers to predict the structural and functional properties of nucleotide sequences at a faster rate than experimental methods can keep up.~\cite{chowdhury_review_2017}

Efficient solutions to this problem can provide essential tools to help biologists research and address pressing issues in their field. For instance, these methods can help biologists track evolutionary processes and better determine relationships among different species~\cite{prjibelski_sequence_2019}. In addition to aiding evolutionary studies, MSA techniques enable researchers to identify regions of similarity, offering insights into the structure, function, and evolutionary relationships of biological sequences (e.g., DNA, RNA, proteins). The sequence alignment process also plays an essential role in genome annotation and gene expression studies.

MSAs are also becoming increasingly significant in fields like personalized medicine and drug design. With their ability to highlight genetic variations, MSAs help predict individual responses to drugs, allowing discoveries into disease susceptibility and potential treatment strategies. Specifically, given their ability to predict ligand selectivity, protein MSAs can have a notable impact on drug development; Drug designers can potentially control and predict the selectivity of drugs, thus improving therapeutic efficacy and minimizing side effects~\cite{gottfries_drug_2006}

Many recent MSA algorithms focus on improving computational efficiency while minimizing alignment error. However, achieving both speed and accuracy remains a challenge, as more accurate alignments often require increased complexity, resulting in longer run times. Unlike pairwise sequence alignment, MSA algorithms must address the complexity of aligning multiple sequences, a computationally intensive problem known to be NP-complete~\cite{wang1994complexity}. The ongoing goal of MSA research is to balance these two aspects, developing methods that compute quickly while producing high-quality alignments.  There are various major types of MSA approaches, each with its own strengths and differences. 

\subsection{Dynamic Programming}

Dynamic programming produces highly precise sequence alignments, but is generally only suitable for small datasets due to its significant computational time and extensive memory requirements. Among the most well-known dynamic programming methods is the Needleman-Wunsch algorithm~\cite{needleman_general_1989}, which performs pairwise global sequence alignments. In contrast, the Smith-Waterman~\cite{smith_identification_1981} algorithm, another prominent approach, is designed for pairwise local sequence alignments, focusing on identifying highly similar regions within sequences. The computational complexity of MSA approaches that use dynamic programming grows exponentially as the number of sequences increases, making them infeasible for large datasets.

While these two algorithms remain relevant in contemporary research, there is a growing shift away from dynamic programming methods as this method is only tractable for pairwise alignments, not multiple sequence alignments. Using the basic concepts of the Needleman-Wunsch and Smith-Waterman algorithms coupled with clever $k$-mer indexing, a much faster approach was developed, BLAST (Basic Local Alignment Search Tool)~\cite{altschul_basic_1990}, in 1990. Multiple versions of BLAST exist today and are still heavily applicable to current biological research, making it one of the few dynamic programming alignment tools used in modern research. Although dynamic programming is not incorporated into many modern MSA approaches, some algorithms based on this technique are still used as benchmarks for the new algorithms currently being developed.

\subsection{Profile HMMs}

Profile-HMM is a type of hidden Markov model (HMM) with a strictly linear, left-to-right structure, widely used in biological sequence analysis~\cite{yoon_hidden_2009}. The main purpose of this approach is to model sequence profiles and infer the existence of events that depend on internal factors that are not always observable. Notable, the sequence profile, an essential input to train an HMM, is itself a multiple sequence alignment. An HMM allows for two processes: an invisible process of hidden states and a visible process of observable symbols (e.g., nucleotides or amino acids).

This structure makes profile-HMMs highly useful for constructing MSAs and allows for more accurate sequence variation analysis. For instance, the Adaptive MSA~\cite{mazooji_fast_2024} algorithm integrates phylogeny-aware profiles and tools like MAGUS + eHMMs~\cite{shen_magusehmms_2022} rely on profile-HMM ensembles. Backbone alignment approaches like UPP~\cite{nguyen_ultra-large_2015}, PASTA~\cite{mirarab_pasta_2015}, and WITCH~\cite{shen_witch_2022} handle sequence-length heterogeneity using profile-HMMs.

However, profile-HMMs assume a linear sequence evolution model, limiting their ability to model more complex evolutionary events like gene duplications or horizontal gene transfer. They also struggle with highly divergent sequences or when there are large gaps in the alignment.

Recent developments in profile-HMMs include the use of deep learning techniques to improve sequence modeling and increase the scalability of MSA algorithms. By combining traditional HMM frameworks with neural networks, researchers are exploring ways to enhance the performance of these models, particularly for large-scale genomic datasets.

\subsection{Progressive Alignment}

Progressive alignment tools are tree-based algorithms that construct a guide tree to determine the branching order for pairwise sequence alignments. The method was first proposed by Hogeweg and Hesper in 1984\cite{hogeweg_alignment_1984}, building on earlier work in sequence alignment and phylogenetic tree construction. This approach is relatively simple, making it computationally efficient. A notable example is T-Coffee (Tree-based Consistency Objective Function For alignmEnt Evaluation)\cite{notredame_t-coffee_2000}, which improves alignment accuracy by clustering sequences initially to identify the most distant ones for alignment. MAGUS\cite{smirnov_magus_2021} uses a Graph Clustering Merger (GCM) to merge constraint alignments using a library of `backbone alignments' that were computed using a `base MSA method', such as MAFFT\cite{katoh_mafft_2002}. With this approach, MAGUS was at least as accurate as PASTA and displayed improvements on the large, challenging datasets.

Nevertheless, progressive alignment methods, including T-Coffee, have several drawbacks. Due to the ``greedy'' nature of the algorithm, errors made early in the alignment process cannot be corrected as more sequences are added to the MSA, potentially lowering the overall quality of the MSA. As a result, more advanced methods, such as iterative refinement approaches like MAFFT or MUSCLE\cite{edgar_muscle_2004}, have been developed to overcome this limitation. Kalign\cite{lassmann_kalign_2020} is another promising MSA approach due to its use of a SIMD (single instruction, multiple data) version of the the bit-parallel Gene Myers algorithm to estimate pairwise distances. Kalign adopted clustal omega's guide tree construction to cluster and later align large datasets of biological sequences. A key limitation of progressive alignment methods is their sensitivity to parameter selection. Similarly, different substitution matrices can affect the alignment of highly divergent sequences, with some matrices favoring more conservative substitutions, while others allow for greater sequence divergence. Because these parameters needed to be optimized for each dataset, their improper selection can lead to significant alignment errors, exacerbating the issues caused by the early stages of progressive alignment, which, in turn, negatively impact the quality of the final MSA\cite{chowdhury_review_2017}.

Progressive alignment methods, including T-Coffee, are widely used in a variety of biological research areas.  However, their limitations become more apparent when aligning highly divergent sequences, where errors in early alignment stages can significantly impact the final MSA. Despite this, progressive alignment methods remain useful in fields such as phylogenetic analysis, where the construction of a reliable guide tree is essential for subsequent sequence alignment. 

\subsection{Genetic Algorithms}

Genetic algorithms (GAs) are optimization methods that use the evolutionary principles of mutation, crossover, and natural selection.  In the context of MSA, closely related sequences typically require fewer modifications to align, enabling the algorithm to focus on refining the best solutions iteratively~\cite{prousalis_survey_2025}.

SAGA~\cite{notredame_saga_1996} was one of †he first multiple sequence alignment approaches built using GAs. The methodology developed in SAGA has since influenced many other more recent algorithms including GA-ACO~\cite{lee_genetic_2008} and MO-SAStrE~\cite{ortuno_optimizing_2013}. GA-ACO incorporates initialization, sexual reproduction operators, mutation operators, fitness evaluation, selection operator, and ant colony optimization to build MSA solutions. MO-SAStrE uses STRIKE score~\cite{kemena_strike_2011}, total conserved (TC) columns, and percentage of non-gaps to further refine alignment qualities. MSAMOGA (Multiple sequence alignment with affine gap by using multi-objective genetic algorithm)\cite{kaya_multiple_2014} was able to find better possible alignments with affine gaps from multiple sequence alignments than competitors such as SAGA and MUSCLE\cite{edgar_muscle_2004}. This algorithm produced high-quality alignments due to its search criteria of similarity of sequences, affine gap penalty and support exhibiting the strength of alignment.

\subsection{Our Approach}

We observed, based on our previous work on CLAM~\cite{ishaq_clustered_2021}, that we had the building blocks for a fast, progressive multiple sequence alignment algorithm.
Building on this work, we introduce MUltiple Sequence Alignment at Scale (\musals{}), which combines CLAM's tree-based hierarchical clustering with star and progressive alignment strategies.

\musals{} is a de novo approach and does not rely on any external information to align the input dataset.
This algorithm uses the Levenshtein distance between sequences to decide the order in which sequences are aligned as we construct the MSA.
Our approach focuses on achieving high-quality alignments while maintaining computational efficiency, even for large datasets.
Using hierarchical clustering, we recursively divide the data set into smaller clusters whose component sequences have small mutual edit-distances and are, thus, easier to align.
We then treat this tree as a guide tree and use a post-order traversal to enable a bottom-up alignment process.

This design uses two alignment strategies; the star alignment method for aligning sequences within clusters and the progressive alignment for merging clusters.
These strategies, applied in distinct stages of the process, balance computational speed with alignment quality.
The result is an approach that efficiently handles large datasets, producing accurate alignments comparable to existing methods while offering significantly faster execution.
\section{Methods}
\label{sec:methods}

\musals{} is a multiple sequence alignment algorithm that first builds a hierarchical cluster tree and then uses it as a guide tree to recursively merge smaller partial alignments into a single MSA for the full dataset.
During clustering, we recursively split the dataset into smaller, internally similar subsets to reduce alignment complexity.
We maintain alignment quality by using Levenshtein distance between sequences instead of approximations such as k-mer profiles or the length of the longest common subsequence.
Having constructed this tree, we perform a post-order traversal to recursively merge alignments from child clusters using their centers to represent their partial alignments.
All alignments in this stage use the Needleman-Wunsh algorithm~\cite{needleman_general_1989} with data-appropriate cost matrices.

\subsection{Clustering}

Clustering is a crucial step in simplifying the computational task of generating an MSA for large datasets.
By dividing the dataset into smaller, more manageable subsets, clustering reduces the computational burden while maintaining the quality of the alignment.
Our clustering approach is hierarchical, adapted from the one used in CLAM~\cite{ishaq_clustered_2021} and CAKES~\cite{prior_let_2024}.
Using Alg.~\ref{alg:methods:cluster-partition}, we build a binary tree  $\mathcal{T}$ of clusters by recursively partitioning the dataset until every leaf cluster contains only one sequence.
We assume that all sequences have been de-duplicated in a pre-processing step before using \musals{}.

\begin{algorithm}[h] 
    \caption{\textsc{Partition}($\mathcal{C}$)} 
    \label{alg:methods:cluster-partition} 
    \begin{algorithmic} 
        \Require $f$, the Levenshtein distance function
        \Require $\mathcal{C}$, a cluster

        \If{$|\mathcal{C}| > 1$}
        \State $c \Leftarrow$ geometric median of $\mathcal{C}$
        \State $l \Leftarrow \argmax f(c, x) \ \forall \ x \in \mathcal{C}$
        \State $r \Leftarrow \argmax f(l, x) \ \forall \ x \in \mathcal{C}$
        \State $\mathcal{L} \Leftarrow \{x \ | \ x \in \mathcal{C} \land f(l, x) \le f(r, x)\}$
        \State $\mathcal{R} \Leftarrow \{x \ | \ x \in \mathcal{C} \land f(r, x) < f(l, x)\}$
        \State \Call{Partition}{$\mathcal{L}$}
        \State \Call{Partition}{$\mathcal{R}$}
        \EndIf
    \end{algorithmic}
\end{algorithm}

Each cluster $\mathcal{C}$ created by this algorithm has a center $c$.
We use the geometric median of the sequences in $\mathcal{C}$ (or a smaller subset for computational efficiency) as the center.
Since $c$ is the sequence which minimizes the sum of distances to all other sequences in $\mathcal{C}$, it serves as the representative for the partial alignment we will later create for $\mathcal{C}$.
We define the radius of $\mathcal{C}$ as the greatest distance from $c$ to any other sequence in $\mathcal{C}$.
Alg.~\ref{alg:methods:cluster-partition} will create progressively smaller children, $\mathcal{L}$ and $\mathcal{R}$, both in terms of cardinality and radius.
This reduces alignment complexity while maintaining quality.

In progressive alignment algorithms, the order in which sequences are aligned greatly impacts the quality of the final alignment. Prioritizing the alignment of similar sequences in this way improves overall alignment quality by reducing unnecessary gap insertions and minimizing alignment errors later on in its process.
In \musals{}, the guide tree is simply the tree resulting from the hierarchical divisive clustering described above.
This approach strengthens the final MSA by helping the algorithm determine the better sequence pairings and alignment order, ultimately enhancing the accuracy and consistency of the alignment.

\subsection{Merging}

\musals{} uses the tree $\mathcal{T}$ from Alg.~\ref{alg:methods:cluster-partition} as the guide tree.
We perform a post-order traversal so we can visit and process child clusters before their parent clusters.
Merging is a recursive procedure outlined as follows:
\begin{itemize}
    \item Base Case: We have a leaf cluster which contains only one sequence. A single sequence is in an MSA by definition.
    \item Recursive Case: We have a parent cluster whose children have already been aligned.
    \begin{enumerate}
        \item From each child, extract its partially aligned center.
        \item Use Needleman-Wunsch to align the two centers and note the indices where gaps were added to each in order to bring them into alignment.
        \item Insert gap columns into the partial alignments of the corresponding child clusters.
        \item The two child clusters have now been aligned to the same width. Join their alignments and use it as the parent cluster's alignment.
    \end{enumerate}
\end{itemize}

This recursive process starts from the leaves of the tree and works its way up to the root, progressively aligning all sequences within the dataset to each other.
By restricting direct alignment operations to cluster centers, the merging process scales linearly with the total number of sequences.

\subsection{Evaluation}

In order to measure the impact \musals{} will have on biological studies, we compared its performance to other well-regarded algorithms.
We evaluate the quality of MSAs produced by \musals{} and its competitors, using a variety of metrics.
For the sake of time and computational resources, we decided to use a random subsample of 10,000 sequences from each alignment to calculate the metrics.

\subsubsection{Percentage of Gaps}

The percentage of gaps measures the ratio of the number of gaps in one aligned sequence to the total length of an aligned sequence.
A low percentage of gaps suggests that the alignment may be of higher quality, as fewer insertions or deletions were required to align the sequences.
However, this metric does not account for mismatches, therefore it cannot fully determine the quality of the MSA.

\subsubsection{P-Score}

To gain a more comprehensive understanding, we analyze the p-score alongside the percentage of gaps.
The p-score evaluates the percentage of mismatched unit pairs while excluding any pairs that include gaps.
By focusing strictly on mismatches, the p-score complements the gap percentage metric, allowing us to assess both aspects of the alignment’s quality.
Better alignments will have a high fraction of matches and, consequently, a low fraction of mismatches.
Therefore a lower p-score represents a higher quality alignment.

\subsubsection{Distance Distortion}

The distance distortion of two sequences is the ratio of the Hamming distance between their aligned versions and the Levenshtein distance between their \textit{unaligned} versions.
We compute the mean distortion over all pairs of sequences.
This metric helps measure the algorithm's ability to preserve the relative dissimilarities between sequences, and of the metric-space of the sequences as a whole.
A lower distance distortion (closer to 1) indicates that the alignment better maintains the original sequence distances, reflecting higher fidelity in preserving evolutionary relationships.

Together, these metrics provide a fuller picture of the MSA’s quality.
The percentage of gaps highlights the extent to which sequences have been modified, the p-score reveals the alignment's accuracy by measuring mismatched pairs, and the distance distortion measures how much the space between sequences has been stretched or compressed.

\subsection{Implementation}

We have implemented \musals{} in the Rust programming language using \texttt{rustc} 1.92.
We leverage the Rayon crate for parallelism and the StringZilla crate for its implementation of Levenshtein distance.
We have our own implementation of the Needleman-Wunsch algorithm.
We use the BLOSUM62 cost-matrix for protein sequences and a cost matrix including the extended-IUPAC alphabet for nucleotides.
While we provide the option of using flat or affine gap penalties, all our experiments use an affine gap penalty where gaps are ten times more expensive to open than they are to extend.

\section{Datasets And Benchmarking}
\label{sec:datasets-and-benchmarks}

We have benchmarked and compared our approach to several different sequence datasets, shown in Table~\ref{tab:datasets:summary}.
Our benchmarks include datasets from three different sources; Greengenes\cite{desantis_greengenes_2006}, the Protein Data Bank (PDB)\cite{berman_protein_2000}, and PFam\cite{bateman_pfam_2002}.
We evaluated our approach and compared it to MAGUS\cite{smirnov_magus_2021}, WITCH\cite{shen_witch_2022}, KAlign\cite{lassmann_kalign_2020}, and FAMSA2\cite{gudys_famsa2_2025},
using the evaluation metrics described in Section~\ref{sec:methods}.

All benchmarks were conducted on an Intel Xeon E5-2690 v4 CPU @ 2.60GHz with 512GB RAM running Manjaro Linux 5.15.164-1.
In all benchmarks, the selected program began with the unaligned sequences and produced an MSA.
The benchmarks aimed to assess multiple aspects of performance, including accuracy, runtime, and scalability.
By combining diverse datasets and rigorous evaluation criteria, we ensured that our method was tested under a wide range of conditions.
These comparisons provide insight into the strengths and limitations of our approach relative to other tools.

\begin{table*}
    \caption{Datasets used in benchmarks.}
    \label{tab:datasets:summary}
    \centering
    \begin{tabular}{|l|r|r|r|r|r|r|}
        \hline
        \multirow{2}{*}{\textbf{Dataset}} & \multirow{2}{*}{\textbf{Cardinality}} & \multicolumn{5}{c|}{\textbf{Lengths}} \\\cline{3-7}
        & & \textbf{Min} & \textbf{Median} & \textbf{Max} & \textbf{Mean} & \textbf{Std. Dev} \\
        \hline
        GreenGenes 12.10 & 1,075,170 & 1,253 & 1,372 & 2,368 & 1,397.43 & 59.36  \\
        \hline
        GreenGenes 13.5  & 1,262,986 & 1,111 & 1,377 & 2,368 & 1,401.06 & 57.14  \\
        \hline
        PDB              & 819,325   & 30    & 224   & 1,000 &   265.54 & 137.43 \\
        \hline
        Pfam 10K Sample & 10,000 & 21 & 241  & 3,629 & 277.61 & 191.89\\ 
        \hline
        Pfam 100K Sample & 100,000 & 8 & 236 & 9,585 & 278.95 & 215.09 \\
        \hline
        Pfam 1M Sample & 1,000,000 & 8 & 255 & 11,910 & 306.86 & 242.16 \\
        \hline
    \end{tabular}
\end{table*}

\subsection{Greengenes}

Greengenes~\cite{desantis_greengenes_2006} is a curated database of 16S rRNA gene sequences widely used in microbial community analyses for taxonomy assignment and phylogenetic studies.
For this study, we used Greengenes 12.10 and Greengenes 13.5.
Greengene 12.10 contains over 1 million and Greengenes 13.5 has over 1.2 million high-quality 16S rRNA sequences representing a wide variety of microbial taxa.
These sequences are curated and aligned to ensure consistency for downstream analyses.
Greengenes serves as a gold standard for evaluating the performance of sequence alignment tools, given its high-quality curation and extensive use in microbial taxonomy studies
Benchmarking against this dataset provides valuable insights into our method's accuracy and scalability.

\subsection{PDB}

The Protein Data Bank (PDB)\cite{berman_protein_2000} dataset is a comprehensive collection of experimentally determined 3D structures of proteins, nucleic acids, and complex assemblies.
For this study, we focused on the protein sequences available in the PDB, which includes a vast range of structures that are essential for various bioinformatics applications, such as sequence alignment, structure-function prediction, and molecular docking studies.
The PDB contains some sequences that are short fragments, and a handful (e.g. titin) that are unusually long; we used only protein sequences with lengths between 30 and 1,000 amino acids, leaving us with just over 800,000 sequences.
While the PDB provides a large number of sequences, its average sequence length is comparatively smaller than those in the Greengenes datasets.

\subsection{PFam}

The Pfam dataset\cite{paysan-lafosse_pfam_2025} is a collection of approximately 81 million protein sequences that are widely used for the analysis of novel genomes and meta genomes.
This dataset was curated with protein families and domains to provide high-quality MSAs and profile HMMs.
Every Pfam family is built from a manually curated seed alignment of representative protein sequencers.
This seed is used to generate an HMM profile to identify homologous sequences in large protein databases.
Pfam provides a diverse collection of protein families with varying sequence lengths, evolutionary divergence, and functional roles.
With this diversity, Pfam is well suited for benchmarking the accuracy, robustness and scalability on multiple sequence alignments on protein data.
Because this dataset is extremely large, the entire dataset was not appropriate to benchmark on several of our comparison algorithms as they would need an unreasonably large amount of memory.
Thus, we created subsets by choosing the first $n$ sequences in order of appearance in the Pfam database, for $n$ of ten thousand, one hundred thousand, and one million.
Note that these samples have mean sequence lengths comparable to the PDB dataset, the maximum lengths of the sequences is extremely high, indicating that there are a small number of very long sequences.
We left these in the sample, in contrast to our subset of PDB, to see how the algorithms would handle such outliers in sequence length.

\section{Results}
\label{sec:results}

The goal of these experiments was to compare our approach to other widely used multiple sequence alignment algorithms to evaluate its relative performance.
We have selected four well-established algorithms for comparasion; MAGUS, WITCH, KAlign and FAMSA2\cite{gudys_famsa2_2025}.
We measured performance across various datasets (Greengenes 12.10, Greengenes 13.5, PDB, and PFam subsets) using the evaluation metrics runtime, alignment width, alignment stretch, percentage of gaps, distance distortion, minimum p distance, mean p distance, and maximum p distance.
We found that several of our comparison algorithms struggled to complete on certain datasets.
MAGUS failed to detect and align subalignments for the protein datasets; it was only able to align the two GreenGenes datasets.
KAlign successfully aligned all Pfam subsets, but encountered difficulties with the PDB dataset because it only supports standard amino acid codes, whereas the dataset contains non-standard residues.

The objective of these benchmarks was to determine whether \musals{} achieves a balance between high alignment quality and improved computational efficiency.

\subsection{Runtime and Scalability}

\musals{} completed alignments on all datasets, and was the only aligner to complete the PDB alignment.
\musals{} was consistently faster than MAGUS, WITCH, and FAMSA, although KAlign was the fastest algorithm on every dataset tested except for Pfam-10k and Pfam-100k.
On Greengeens 13.5, \musals{} was approximately 15 times faster than WITCH, 5 times faster than FAMSA and 4.5 times faster than MAGUS
As dataset size increases from the PFam subsamples, \musals{} exhibited a gradual increase in runtime, showing a near-linear scaling with the number of sequences.
For the Pfam subsets, \musals{}' runtime increased from 18 seconds for the 10k to 365 seconds for the 100k and 2,348 seconds for the 1M. 
Runtime results are visualized in Fig.~\ref{fig:runtime}.

\begin{table*}[t]
\FloatBarrier
\caption{Quality and runtime comparison on benchmark datasets. MAGUS had difficulties running on protein sequences and therefore failed before getting an alignment for the PDB and the Pfam datasets.}
\label{tab:greengenes_comparison}
\centering
\begin{tabular}{l l r r r r r r r r}
\toprule
\textbf{Dataset} &
\textbf{Alg.} &
\textbf{Time (s)} &
\textbf{Width} &
\textbf{Stretch} &
\textbf{\% Gaps} &
\textbf{Distortion} &
\textbf{P$_{min}$} &
\textbf{P$_{avg}$} &
\textbf{P$_{max}$} \\
\midrule

\multirow{4}{*}{Greengenes 12.10}
 & MAGUS  & $3.85\mathrm{e}{5}$ & 93,008 & 39.28 & 98.50 & 1.1822 & 0 & 0.0032 & 0.0064 \\
 & WITCH  & $1.16\mathrm{e}{6}$ & 35,986 & 15.20 & 96.12 & 1.1367 & 0 & 0.0084 & 0.0164 \\
 & KAlign & $2.96\mathrm{e}{3}$ & 40,565 & 17.13 & 96.56 & 1.1361 & 0 & 0.0074 & 0.0142 \\
 & FAMSA  & -- & -- & -- & -- & -- & -- & -- & -- \\
 & \musals{} & $9.73\mathrm{e}{4}$ & 2,789 & 1.18 & 49.93 & 5.3804 & 0 & 0.5818 & 0.7020 \\
\midrule

\multirow{4}{*}{Greengenes 13.5}
 & MAGUS  & $4.47\mathrm{e}{5}$ & 79,177 & 33.44 & 98.23 & 1.1896 & 0 & 0.0038 & 0.0082 \\
 & WITCH  & $1.57\mathrm{e}{6}$ & 55,218 & 23.32 & 97.47 & 1.1397 & 0 & 0.0054 & 0.0120 \\
 & KAlign & $3.24\mathrm{e}{3}$ & 54,349 & 22.95 & 97.42 & 1.1392 & 0 & 0.0055 & 0.0111 \\
 & FAMSA  & $5.24\mathrm{e}{5}$ & 84,375 & 35.63 & 98.34 & 1.1423 & 0 & 0.0035 & 0.0071 \\
 & \musals{} & $1.07\mathrm{e}{5}$ & 3,070 & 1.30 & 54.36 & 5.8532 & 0 & 0.5672 & 0.7873 \\
\midrule

\multirow{4}{*}{PDB}
 & MAGUS  & -- & -- & -- & -- & -- & -- & -- & -- \\
 & WITCH  & -- & -- & -- & -- & -- & -- & -- & -- \\
 & KAlign & -- & -- & -- & -- & -- & -- & -- & -- \\
 & FAMSA  & -- & -- & -- & -- & -- & -- & -- & -- \\
 & \musals{} & $3.14\mathrm{e}{3}$ & 1,087 & 1.09 & 75.57 & 1.2182 & 0 & 0.3160 & 0.9522 \\
\midrule

\multirow{4}{*}{PFam 10k}
 & MAGUS  & -- & -- & -- & -- & -- & -- & -- & -- \\
 & WITCH  & -- & -- & -- & -- & -- & -- & -- & -- \\
 & KAlign & $3.20\mathrm{e}{1}$ & 4,243 & 1.17 & 93.42 & 1.4128 & 0.0016 & 0.0274 & 0.2800 \\
 & FAMSA  & $4.15\mathrm{e}{4}$ & 9,178 & 2.53 & 96.89 & 1.2611 & 0 & 0.0162 & 0.1424 \\
 & \musals{} & $1.80\mathrm{e}{1}$ & 3,629 & 1.00 & 92.35 & 1.2137 & 0.0061 & 0.1012 & 0.5767 \\
\midrule

\multirow{4}{*}{PFam 100k}
 & MAGUS  & --   & -- & -- & -- & -- & -- & -- & -- \\
 & WITCH  & --   & -- & -- & -- & -- & -- & -- & -- \\
 & KAlign & $3.65\mathrm{e}{2}$ & 10,047 & 1.05 & 97.25 & 1.6300 &  0 & 0.0057 & 0.0815 \\
 & FAMSA  & --   & -- & -- & -- & -- & -- & -- & -- \\
 & \musals{} & $2.48\mathrm{e}{2}$ & 9,585 & 1.00 & 97.09 & 1.2128 & 0.0034 & 0.0397 & 0.3327 \\
\midrule

\multirow{4}{*}{PFam 1M}
 & MAGUS  & -- & -- & -- & -- & -- & -- & -- \\
 & WITCH  & -- & -- & -- & -- & -- & -- & -- \\
 & KAlign & $2.35\mathrm{e}{3}$ & 23,120 & 1.94 & 98.67 & 1.7753 & 0 & 0.0012 & 0.0487 \\
 & FAMSA  & -- & -- & -- & -- & -- & -- & -- \\
 & \musals{} & $3.39\mathrm{e}{3}$ & 11,924 & 1.00 & 97.66 & 1.2181 & 0.0020 & 0.0349 & 0.2093 \\
\midrule
\end{tabular}
\end{table*}

\subsection{Alignment Accuracy}

As shown in Table~\ref{tab:greengenes_comparison}, \musals{} constructed a much more compact alignment than all competing methods. For GreenGenes 13.5, \musals{} achieved a MSA with a width of only 3,070 while the MSAs computed by MAGUS, WITCH, KAlign, and FAMSA had widths of 79,177, 55,218, 54,349, and 84,375, respectively.
This compactness essentially equates to fewer gaps and thus more mismatches. The gap percentage is visualized in Fig.~\ref{fig:gaps}.

In terms of distance distortion and average P-score, \musals{} achieves the highest (worst) scores on the Greengenes datasets, which were the only datasets on which all other aligners could complete their alignments.
However, on the Pfam datasets, where Kalign also succeeded, \musals{} outperforms Kalign in terms of distance distortion, achieving scores of 1.2137 vs. 1.4128 on Pfam-10k, 1.2128 vs. 1.6300 on Pfam-100k, and 1.2181 vs. 1.7753 on Pfam-1M.
It also outperforms FAMSA (1.2611) on the one dataset, Pfam-10k, that aligner was able to complete.
Kalign does achieve better average P-scores on these datasets.

\definecolor{\musals{}}{RGB}{31,119,180}
\definecolor{MAGUS}{RGB}{255,127,14}
\definecolor{WITCH}{RGB}{44,160,44}
\definecolor{KAlign}{RGB}{214,39,40}
\definecolor{FAMSA}{RGB}{148,103,189}

\begin{figure*}[t]
\label{fig:runtime}
\centering
\pgfplotsset{
/pgfplots/my legend/.style={
legend image code/.code={
\draw[thick,black](-0.05cm,0cm) -- (0.3cm,0cm);%
   }
  }
}

\begin{tikzpicture}
    \begin{axis}[
        width = \textwidth,
        height = 14cm,
        major x tick style = transparent,
        ybar,
        bar width=8pt,
        ymajorgrids = true,
        ylabel = {Run time speed},
        symbolic x coords={GG12\_10, GG13\_5,PDB, PFam\_10K, PFam\_100K, PFam\_1M},
        xtick = data,
        scaled y ticks = false,
        enlarge x limits=0.1,
        ymode=log,
        log basis y=10,
        ymin=1, 
        ytick={1,10,100,1e3,1e4,1e5,1e6,1e7},
        ylabel={Run time (seconds)},
        legend cell align=left,
        legend style={
                at={(1,1.05)},
                anchor=south east,
                column sep=1ex
        },
        extra y ticks = 0.4,
        extra y tick labels={},
        extra y tick style={grid=major,major grid style={thick,draw=black}}
    ]
        \addplot[style={MAGUS,fill=MAGUS,mark=none}]
            coordinates {(GG12\_10, 384860) (GG13\_5, 447146) (PDB, 1) (PFam\_10K, 1) (PFam\_100K, 1) (PFam\_1M, 1)};

        \addplot[style={WITCH,fill=WITCH,mark=none}]
             coordinates {(GG12\_10, 1164760) (GG13\_5, 1566556) (PDB, 1) (PFam\_10K, 1) (PFam\_100K, 1) (PFam\_1M, 1)};
             
        \addplot[style={KAlign,fill=KAlign,mark=none}]
             coordinates {(GG12\_10, 2957) (GG13\_5, 3241) (PDB, 1) (PFam\_10K, 32) (PFam\_100K, 365) (PFam\_1M, 2348)};

        \addplot[style={FAMSA,fill=FAMSA,mark=none}]
             coordinates {(GG12\_10, 1) (GG13\_5, 524116) (PDB, 1) (PFam\_10K, 41534) (PFam\_100K, 1) (PFam\_1M, 1)};

        \addplot[style={\musals{},fill=\musals{},mark=none}]
             coordinates {(GG12\_10, 97326) (GG13\_5, 106944) (PDB, 3137) (PFam\_10K, 18) (PFam\_100K, 248) (PFam\_1M, 3390)};

        \legend{MAGUS, WITCH, KAlign, FAMSA, MuSAlS}
        \addlegendimage{my legend}
    \end{axis}
\end{tikzpicture}
\caption{Runtime comparison of MSA algorithms (log scale) with grouped bars.}
\label{fig:runtime}
\end{figure*}
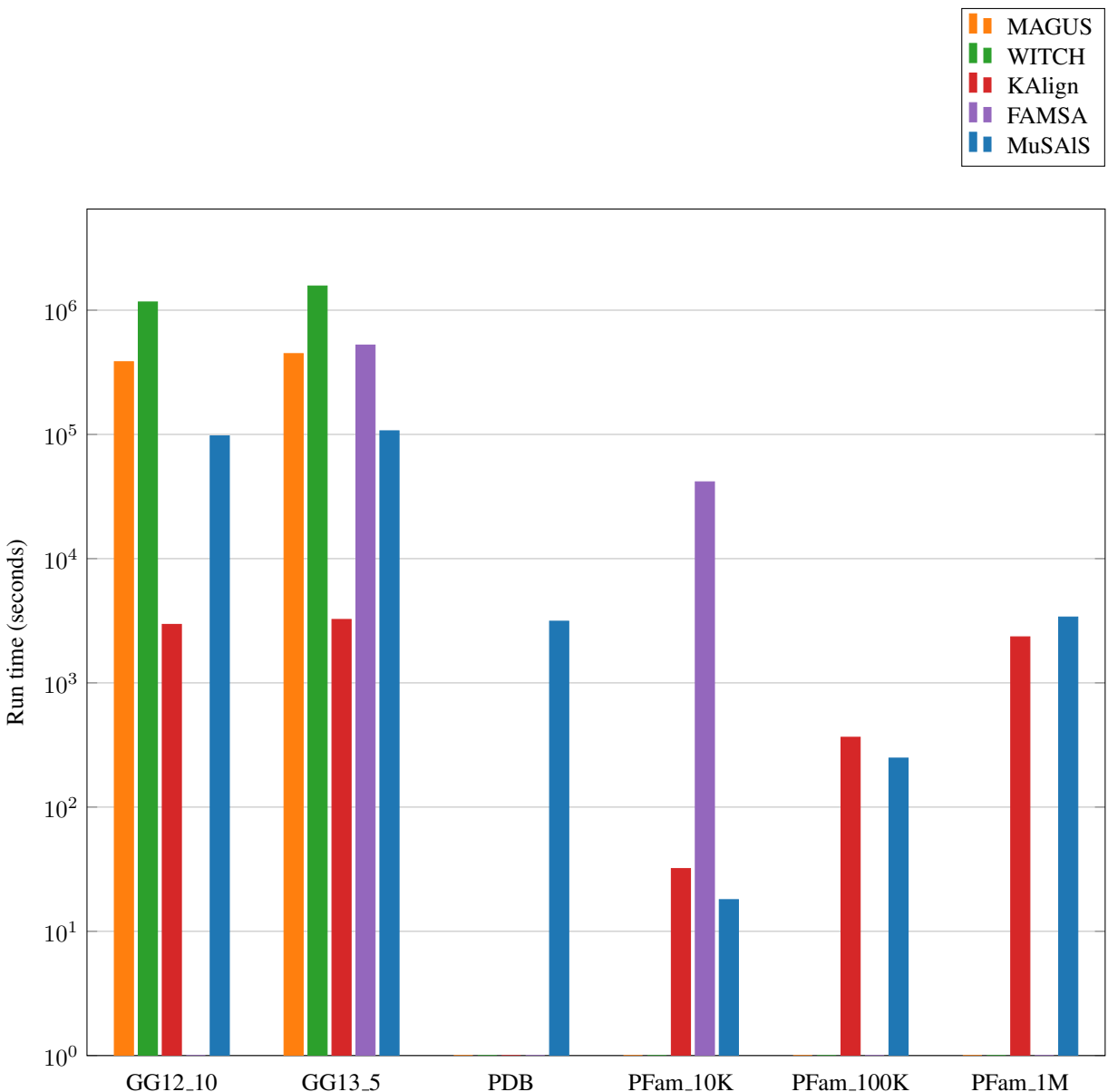

\begin{figure*}[t]
\centering
\pgfplotsset{
/pgfplots/my legend/.style={
legend image code/.code={
\draw[thick,black](-0.05cm,0cm) -- (0.3cm,0cm);%
   }
  }
}

\begin{tikzpicture}
    \begin{axis}[
        width = \textwidth,
        height = 14cm,
        major x tick style = transparent,
        ybar,
        bar width=10pt,
        ymajorgrids = true,
        ylabel = {Percentage of Gaps},
        symbolic x coords={GG12\_10, GG13\_5,PDB, PFam\_10K, PFam\_100K, PFam\_1M},
        xtick = data,
        scaled y ticks = false,
        enlarge x limits=0.1,
        ymin=0,
        legend cell align=left,
        legend style={
                at={(1,1.05)},
                anchor=south east,
                column sep=1ex
        },
        extra y ticks = 0.4,
        extra y tick labels={},
        extra y tick style={grid=major,major grid style={thick,draw=black}}
    ]
        \addplot[style={MAGUS,fill=MAGUS,mark=none}]
            coordinates {(GG12\_10, 98.5) (GG13\_5, 98.23) (PDB, 0) (PFam\_10K, 0) (PFam\_100K, 0) (PFam\_1M, 0)};

        \addplot[style={WITCH,fill=WITCH,mark=none}]
             coordinates {(GG12\_10, 96.12) (GG13\_5, 97.46) (PDB, 0) (PFam\_10K, 0) (PFam\_100K, 0) (PFam\_1M, 0)};
             
        \addplot[style={KAlign,fill=KAlign,mark=none}]
             coordinates {(GG12\_10, 96.56) (GG13\_5, 97.42) (PDB, 0) (PFam\_10K, 0) (PFam\_100K, 0) (PFam\_1M, 0)};

        \addplot[style={FAMSA,fill=FAMSA,mark=none}]
             coordinates {(GG12\_10, 0) (GG13\_5, 98.34) (PDB, 0) (PFam\_10K, 96.98) (PFam\_100K, 0) (PFam\_1M, 0)};

        \addplot[style={\musals{},fill=\musals{},mark=none}]
             coordinates {(GG12\_10, 49.93) (GG13\_5, 
             54.36) (PDB, 75.57) (PFam\_10K, 92.35) (PFam\_100K, 97.09) (PFam\_1M, 97.66)};

        \legend{MAGUS, WITCH, KAlign, FAMSA, \musals{}}
        \addlegendimage{my legend}
    \end{axis}
\end{tikzpicture}
\caption{Percentage of gaps comparison of MSA algorithms. A lower percentage of gaps represents a tighter alignment}
\label{fig:gaps}
\end{figure*}
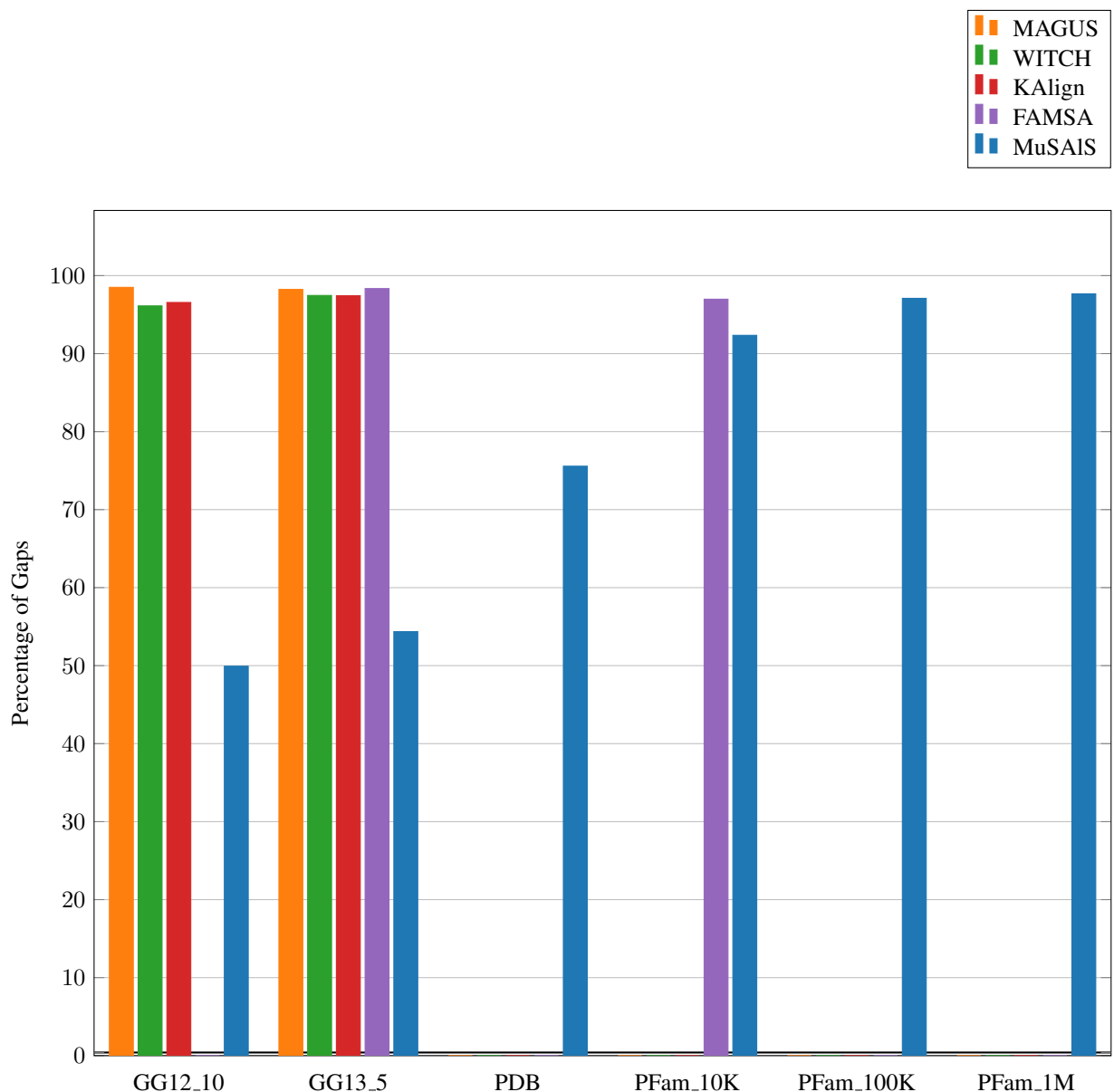

\section{Discussion and Future Work}
\label{sec:discussion-and-future-work}

There have been many different MSA algorithms that have been developed over the years, each with their own strengths and weaknesses.
\musals{}, similar to the other MSA tools, has room for improvements despite its ability to quickly and accurately align a large number of sequences.
One distinguishing feature of \musals{} is its de novo alignment approach, which does not rely on external information such as evolutionary or structural data.
While this simplicity can be advantageous in certain contexts, it also means that other MSA tools that incorporate such external information can produce higher-quality MSAs in scenarios where such data is available.

\musals{} tends to produce alignments that are an order of magnitude more compact than the other aligners to which we compared it. This, of course, represents a tradeoff; \musals{} produces alignments with more mismatches than others. However, on the protein datasets, it produced less distance-distorted alignments than the two other tools we were able to run; this may be due to the mismatch penalty being relatively simple for nucleotides but based on evolutionary information (BLOSUM62) for amino acids. The bias towards compactness may also be due in part to the use of affine gap penalties; future experimentation could compare the results using flat gap penalties. The tension between compact and consistent alignments (that is, those with few mismatches), in the absence of Pareto-optimal results, suggests further exploration of these tradeoffs. Namely, Is there a single combined evaluation approach that could be used?

However, in the absence of external evolutionary information, such as in the analysis of non-model organisms or datasets where multiple genomes exhibit only small differences, \musals{} proves to be a very fast and scalable de novo aligner.
Its design effectively addresses the scalability challenge, allowing it to align large datasets with high efficiency.
This makes \musals{} particularly well-suited for high-throughput sequencing applications, where rapid and accurate alignment of numerous sequences is critical.
One future improvement for \musals{} would be to enable online updates --- that is, to allow new sequences to be added to the dataset, dynamically updating the clusters, the guide tree, and the MSA.

A significant limitation at present is that while \musals{} demonstrably scales well with the \emph{number} of sequences, it is still limited by its quadratic time and space complexity in the \emph{length} of the sequences.
\musals{} inherits these limitations from the Levenshtein and Needleman-Wunsch algorithms.
While some of the aligned sequences are thousands of nucleotides or amino acids in length, this is still effectively a gene-scale alignment.
Aligning entire chromosomes or bacterial genomes is not currently feasible with \musals{}.
However, heuristics to improve the performance of our Needleman-Wunsch implementation, such as $k$-mer anchors to split sequences into smaller alignable sub-sequences, would allow \musals{} to scale to large numbers of very wide bacterial or chromosomal sequences.
By addressing these limitations, \musals{} could extend its applicability to a wider range of data sets and research contexts.

\musals{}' efficiency and ease of use make it a valuable tool for studying genomes and evolutionary relationships.
Its scalability and rapid processing make it especially useful for applications such as metagenomics, population genetics, and phylogenetics.
By addressing current limitations and building on its strengths, \musals{} has the potential to inspire the development of new, scalable alignment algorithms, contributing to the growing demand for efficient and reliable computational biology tools in genomic research.


\bibliographystyle{plain}
\bibliography{references.bib}  

\end{document}